\documentstyle[12pt,epsfig]{article}
\newcommand{\beq}{\begin{equation}}
\newcommand{\eeq}{\end{equation}}
\newcommand{\beqa}{\begin{eqnarray}}
\newcommand{\eeqa}{\end{eqnarray}}
\newcommand{\beqan}{\begin{eqnarray*}}
\newcommand{\eeqan}{\end{eqnarray*}}
\newcommand{\ba}{\begin{array}}
\newcommand{\ea}{\end{array}}
\newcommand{\ben}{\begin{enumerate}}
\newcommand{\een}{\end{enumerate}}
\newcommand{\bfl}{\begin{flushleft}}
\newcommand{\efl}{\end{flushleft}}
\newcommand{\btab}{\begin{tabular}}
\newcommand{\etab}{\end{tabular}}
\newcommand{\bit}{\begin{itemize}}
\newcommand{\eit}{\end{itemize}}
\newcommand{\bdes}{\begin{description}}
\newcommand{\edes}{\end{description}}

\newcommand{\ol}{\overline}
\newcommand{\ra}{\rightarrow}

\newcommand{\ve}{\varepsilon}

\newcommand{\dfrac}{\displaystyle \frac}

\begin{document}
\parskip=4pt plus 1pt  %inserts small space between paragraphs
\begin{titlepage}
\begin{flushright}
UWThPh-1996-62\\
Oct. 1996
\end{flushright}
\vspace{2cm}
\begin{center}
{\Large \bf
STATUS OF CHIRAL PERTURBATION THEORY} \\[30pt]
{\bf Gerhard Ecker} \\[10pt]
Institut f\"ur Theoretische Physik, Universit\"at Wien \\[5pt]
Boltzmanngasse 5, A--1090 Wien, Austria \\[50pt]
{\bf ABSTRACT} \\[10pt]
\end{center}
\noindent
A survey is made of semileptonic and nonleptonic kaon decays in the 
framework of chiral perturbation theory. The emphasis is on what has
been done rather than how it was done. The theoretical predictions are
compared with available experimental results.
 
\vfill
\begin{center}
Talk given at the \\[5pt]
Workshop on Heavy Quarks at Fixed Target\\[5pt] 
Rheinfels Castle, St. Goar, Germany\\[5pt]
Oct. 3 - 6, 1996 \\[5pt]
To appear in the Proceedings
\end{center}
\vfill

\end{titlepage}
\setcounter{page}{1}
%\newpage
%
\section{Physics at Low Energies}
By decision of the Organizing Committee of this meeting, the strange
quark was declared to be a heavy quark. Although we have benefitted from
this decision, the strange quark is really a light quark on most
other accounts. Of course, all quarks except for the top quark are
light compared to the natural scale $M_W$ of the Standard Model.
However, there is an important practical distinction between bottom
and charm quarks on one side and up, down and strange quarks on
the other side.

As we come down from the ``fundamental" scale $M_W$ to
lower energies, we can rely on perturbative QCD for the operator
product expansion to describe physics at energies down to 
about $m_c$ \cite{Buchalla}. At this scale, the Lagrangian of the 
Standard Model has broken
up into different pieces, such as the strangeness changing Lagrangian
with $|\Delta S|=1$ relevant for kaon physics. The degrees of freedom
in those Lagrangians are the gluons and the quarks with masses
below 1~GeV. From here on, the picture changes drastically as far
as the theoretical framework is concerned. Because of confinement,
it does not make sense to use perturbative QCD to describe the interactions 
of ``light" quarks at energies below 1~GeV. 

Among the many models and methods that have been employed to describe
physics at low energies, two of them have the best theoretical credits
by far: lattice gauge theories \cite{Guesken} and chiral perturbation
theory \cite{Wein79,GL84,GL85a,Leut94} (CHPT). The strategy of CHPT is 
that of an effective
field theory for the actually observed degrees of freedom, i.e. for
the hadrons (and leptons). CHPT uses only the symmetries of the Standard 
Model to construct 
the effective field theory in the nonperturbative domain. The advantage
of such an approach is its generality: the predictions of CHPT are rigorous
predictions of the Standard Model. The drawback is that this effective
field theory has a score of a priori undetermined coupling constants (often
called low--energy constants) that are not constrained by
the symmetries. As is true for any effective field theory, these 
low--energy constants are remnants of the ``short--distance" structure.
The notion ``short distances" encompasses all degrees of freedom
that are not included as explicit fields in the effective Lagrangians.
For kaon physics, only the pseudoscalar meson octet is contained in
the CHPT Lagrangians. All other effects at higher scales such as
meson resonances or short--distance effects in the usual terminology like
CP violation are incorporated in the low--energy constants. Let me note
at this point that CP violation will not be discussed in this survey.

An essential ingredient of CHPT is the spontaneously broken chiral symmetry,
an approximate symmetry of the Standard Model for light quarks. The
structure of the CHPT Lagrangians is constrained by this symmetry that
gives rise to a systematic low--energy expansion. The relevant expansion
parameter is
\beq
\dfrac{p^2}{16 \pi^2 F_\pi^2}=\dfrac{p^2}{M_K^2}.
\dfrac{M_K^2}{16 \pi^2 F_\pi^2}=0.18~\dfrac{p^2}{M_K^2}
\eeq
where $p$ is a typical momentum and $F_\pi=92.4$~MeV is the pion decay
constant. Therefore, higher--order corrections in CHPT amplitudes for kaon 
decays are naturally of the order of $20\%$. This is not a very small 
parameter, but it is small enough to talk of a perturbative 
expansion unlike for the strong coupling constant in this energy range.

The status of CHPT is discussed in several recent reviews \cite
{PPNP95,Pich95,EdR95,BKM95}. For kaon decays in general and for the
chiral description in particular, the standard reference is the Second
DA$\Phi$NE Handbook of Physics \cite{DAF2}. In fact, most of what I
am going to cover here can already be found there. More recent accounts
can be found in the Proceedings of the Workshop on $K$ Physics held
at Orsay this spring \cite{Orsay}.

\section{Semileptonic $K$ Decays}
All semileptonic $K$ decays that can be measured in the foreseeable
future have been calculated at one--loop level. In the standard
CHPT terminology, this corresponds to $O(p^4)$ for amplitudes without
an $\ve$ tensor and $O(p^6)$ for those with an $\ve$ tensor. In some cases,
higher--order corrections have been at least partly included with
the help of dispersion relations. For semileptonic decays,
the Second DA$\Phi$NE Handbook \cite{DAF2} is still up--to--date. I will
therefore restrict myself to a few illustrative examples.

\renewcommand{\arraystretch}{1.1}
\begin{table}[t]
\begin{center}
\caption{\it Phenomenological values and source for the renormalized coupling
constants $L^r_i(M_\rho)$. The errors include estimates of the effect
of higher--order corrections.}\label{tab:Li}
\vspace{.5cm}
\begin{tabular}{|c||r|l|}  \hline
i & $L^r_i(M_\rho) \times 10^3$ & source  \\ \hline
  1  & 0.4 $\pm$ 0.3 & $K_{e4},\pi\pi\rightarrow\pi\pi$   \\
  2  & 1.35 $\pm$ 0.3 &  $K_{e4},\pi\pi\rightarrow\pi\pi$  \\
  3  & $-$3.5 $\pm$ 1.1 &$K_{e4},\pi\pi\rightarrow\pi\pi$   \\
  4  & $-$0.3 $\pm$ 0.5 & Zweig rule  \\
  5  & 1.4 $\pm$ 0.5  & $F_K:F_\pi$  \\
  6  & $-$0.2 $\pm$ 0.3 & Zweig rule  \\
  7  & $-$0.4 $\pm$ 0.2 &Gell-Mann--Okubo,$L_5,L_8$     \\
  8  & 0.9 $\pm$ 0.3 & \small{$M_{K^0}-M_{K^+},L_5,$} \\
     &               &   \small{ $(2m_s-m_u-m_d):(m_d-m_u)$}  \\
 9  & 6.9 $\pm$ 0.7 & $\langle r^2\rangle^\pi_V$  \\
 10  & $-$5.5 $\pm$ 0.7& $\pi \rightarrow e \nu\gamma$   \\
\hline
\end{tabular}
\end{center}
\end{table}

At lowest order in the low--energy expansion, the chiral Lagrangian
for the strong, electromagnetic and semileptonic weak interactions
contains a single parameter $F$, equal to $F_\pi=F_K$ at this level,
and the meson masses. At the next order, $O(p^4)$,
there are 10 new low--energy constants \cite{GL85a} 
$L_1$,\dots,$L_{10}$. Since they have all been determined 
phenomenologically, we have a completely predictive scheme to this order.
In Table \ref{tab:Li}, the current status of these low--energy constants
is summarized.

As a first example, we consider the decays $K_{l2ll}$, i.e. $K_{l2}$
decays with a virtual photon producing a lepton pair. There are
four form factors characterizing the decay amplitudes. The three axial
ones dominate the rates. In Table \ref{tab:Kl2},
the chiral predictions \cite{BEG93} are compared with available experimental
results. The decay mode with an electron neutrino is especially interesting
because practically the whole amplitude is generated at $O(p^4)$
due to the helicity suppression of the lowest--order amplitude
(Bremsstrahlung). In both channels where events have been found the
effects of $O(p^4)$ are definitely seen.

\begin{table}
\begin{center}
\caption{\it Branching ratios for $K_{l2ll}$ from theory \protect\cite{BEG93}
and experiment \protect\cite{Diamant76,Atiya89}. For the modes with an $e^+e^-$
pair, a cut $m_{e^+e^-}\geq$~140 MeV has been applied.}\label{tab:Kl2}
\vspace{.5cm}
\begin{tabular}{|c||c|c|c|}
\hline
 & $K^+\to\mu^+ \nu_\mu e^+ e^- $&$ K^+\to e^+\nu_e e^+e^- $&$
K^+\to\mu^+\nu_\mu \mu^+\mu^-$\\
\hline
\rule{0cm}{6mm} tree & $5.0\cdot 10^{-8} $&$2.1\cdot10^{-12}$&$3.8\cdot10^{-9}$\\
1-loop & $8.5\cdot10^{-8}$&$3.4\cdot10^{-8}$&$1.35\cdot10^{-8}$\\
experiment & $(1.23\pm0.32)\cdot
10^{-7}$&$(2.8^{+2.8}_{-1.4})\cdot 10^{-8}$&
$\leq 4.1\cdot10^{-7}$ \\
\hline
\end{tabular}
\end{center}
\end{table}

The second class of decay modes I want to mention are radiative $K_{l3}$
decays. The theoretical analysis \cite{BEG93} involves
altogether ten form factors two of which appear also in the non--radiative
decays. The final conclusion after quite some work is that the
effects of $O(p^4)$ are relatively small: the amplitudes are still dominated
by Bremsstrahlung. Let me emphasize that this is a definite
prediction rather than an unfortunate mishap. As we saw in the previous
case and as we shall see again for $K_{l4}$ decays, the corrections of 
$O(p^4)$ are by no means negligibly small in general. Instead of a 
comprehensive comparison with experiment (for the status as of 1995 
see Ref.~\cite{BCEG95}), I concentrate in Table \ref{tab:Kl3g} on the decays 
$K_L\to \pi^\pm e^\mp \nu_e \gamma$
where a new experimental result \cite{Leber96} has become available.
The chiral prediction that the decay is dominated by Bremsstrahlung is 
supported by the data.
\begin{table}
\begin{center}
\caption{\it Branching ratio for $K_L\to \pi^\pm e^\mp \nu_e \gamma$
from theory \protect\cite{BEG93} and experiment \protect\cite{Leber96}. Cuts
on the photon energy $E_\gamma\geq$ 30 MeV and on the electron--photon
opening angle $\Theta_{e\gamma}\geq$ $20^o$ have been applied.}
\label{tab:Kl3g}
\vspace{.5cm}
\begin{tabular}{|c||c|}
\hline
 & BR($K_L\to \pi^\pm e^\mp \nu_e \gamma$)\\
\hline
Bremsstrahlung & $3.6\cdot 10^{-3}$ \\
$O(p^4)$ ($L_i$ only) & $4.0\cdot10^{-3}$\\
$O(p^4)$ total ($L_i$ and loops) & $3.8\cdot10^{-3}$\\
experiment (NA31) & $(3.61\pm0.14\pm^{0.21}_{0.15})\cdot 10^{-3}$ \\
\hline
\end{tabular}
\end{center}
\end{table}

As a final example of semileptonic decays let me turn to $K_{l4}$
decays. As for $K_{l2ll}$, there are three axial and one vector form factor. 
Two of the axial form factors dominate the amplitudes. In addition
to the $O(p^4)$ calculation \cite{Bij90}, the dominant higher--order
effects were estimated using dispersion relations \cite{BCG94}. In
this case, the corrections of the leading current algebra amplitudes of 
$O(p^2)$ are large. In Table \ref{tab:Kl4} taken from the talk of Bijnens
at the Orsay Workshop \cite{Orsay}, the chiral predictions \cite{BCG94} for the 
various decay widths are confronted with experiment. The
channel $K^+\to \pi^+\pi^- e^+ \nu_e$ with the highest statistics \cite{
Rosselet77} was used to extract the three low--energy constants
$L_1$, $L_2$ and $L_3$ together with information from $\pi\pi$
scattering (cf. Table \ref{tab:Li}). The agreement with experiment in
the remaining channels is quite impressive although the statistics is
limited. Note the big corrections in going from tree level to $O(p^4)$
and the still sizable higher--order corrections (third line of theoretical
predictions in Table \ref{tab:Kl4} denoted ``full").
\begin{table}
\begin{center}
\caption{\it Predictions for the various $K_{l4}$ decay 
widths \protect\cite{BCG94}. The last two columns are normalized to 
$K_L$ decays. Full includes estimates of higher--order corrections
beyond $O(p^4)$. Errors are 
in brackets and all values are in $s^{-1}$.}
\label{tab:Kl4}
\vspace{.5cm}
\begin{tabular}{|c|cccccc|}
\hline
$\pi\pi$ charge & $+-$ & $00$ & $+-$ & $00$ & $0-$ & $0-$ \\
leptons & $e^+\nu$ & $e^+\nu$ & $\mu^+\nu$ & $\mu^+\nu$ &$e^+\nu$ &
$\mu^+\nu$\\
\hline
tree & 1297 & 683 & 155 &  102 & 561 & 55\\
$p^4$ & 2447 & 1301 & 288 & 189 & 953 & 94 \\
full & input & 1625(90) & 333(15) & 225(11) & 917(170) & 88(22)\\
\hline
exp. & 3160(140) & 1700(320) & 1130(730) & $-$ & 998(80) & $-$\\
\hline
\end{tabular}
\end{center}
\end{table}

\begin{figure}[t]
\unitlength1cm
\begin{picture}(2,1) \end{picture}
\epsfysize=8cm
\epsffile{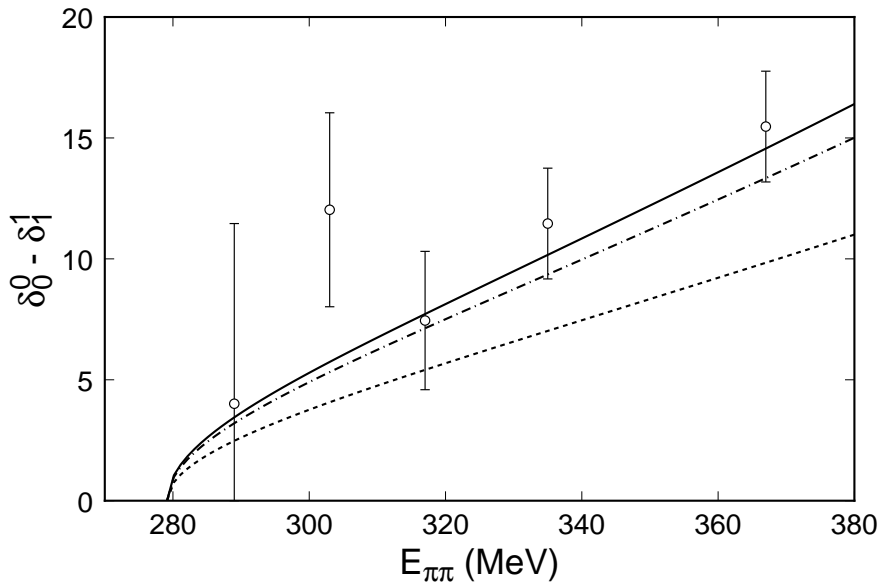}
\caption{\it The phase shift difference $\delta_0^0-\delta_1^1$ (in degrees) 
as a function of the center--of--mass energy of the two incoming pions. The 
dotted (dash--dotted) line displays the tree (one--loop) approximation,
whereas the solid line denotes the two--loop result \protect\cite{BCEGS96}. 
The data are from Rosselet et al. \protect\cite{Rosselet77}.} \label{fig:pipi}
\end{figure}

The rates are mainly determined by the real part of the form factors.
Through the imaginary parts, $K_{l4}$ decays also allow for
accurate measurements of some of the $\pi\pi$ scattering phase shifts.
The present experimental status \cite{Rosselet77} is shown in Fig. 
\ref{fig:pipi} together with
the theoretical predictions up to $O(p^6)$ (two--loop level) for the 
difference between the $I=l=0$ and the $I=l=1$ phase shifts \cite{BCEGS96}. 
We are looking forward to precision data on $K_{e4}$ from DA$\Phi$NE and 
other kaon facilities to test CHPT to $O(p^6)$. This is a calculation
based on chiral $SU(2)$ (pions only) where the natural expansion parameter
is much smaller than for kaon decays, at least near threshold.

\section{Nonleptonic $K$ Decays}
Already at the level of the operator product expansion, the
semileptonic and nonleptonic weak decays are described by different
Lagrangians. The same holds for the effective description
of CHPT. The nonleptonic weak interactions of kaons require an additional
chiral Lagrangian with low--energy constants that have a priori
nothing to do with the strong constants $L_i$ in Table \ref{tab:Li}.

At lowest order, again $O(p^2)$, the chiral Lagrangian for 
$|\Delta S|=1$ nonleptonic weak interactions is characterized by two
coupling constants $G_8$ and $G_{27}$, responsible for the octet and
the 27--plet part of the effective Hamiltonian. The dominant decay
modes  $K\to 2 \pi,\,3 \pi$ are determined by these constants at
lowest order (current algebra level). From $K\to 2 \pi$ decays one
extracts
\beq
|G_8| = 9\cdot 10^{-6} {\rm GeV}^{-2}~,\qquad
G_{27}/G_8 = 1/18~. \label{G827}
\eeq
The small ratio between $G_{27}$ and $G_8$ is a manifestation, but
of course not an explanation of the $\Delta I=1/2$ rule. For CHPT, this
small ratio is input that allows to neglect the 27--plet
contribution in most cases where the octet also contributes. With the
values (\ref{G827}) one can predict 7 measurable quantities in $K\to
3 \pi$ decays (amplitude and slope parameters). The conclusion has been
known for many years: the agreement is qualitative only and
there are sizable deviations on the order of 20 -- 30 $\%$ in amplitude, 
precisely of the order expected in CHPT.

Before we move on to the next order in the chiral expansion, we
can ask ourselves whether there are other channels for which predictions
can be made at lowest order. I am not aware of another framework where
one could prove the following statement as easily as in CHPT: there
is no additional information in nonleptonic kaon amplitudes at
lowest order in the momentum expansion beyond Bremsstrahlung which is 
of course determined by
the non--radiative $K\to 2 \pi,\,\,3\pi$ decays. In other words, most 
of the interesting physics in nonleptonic kaon decays starts at $O(p^4)$ 
only.

At next--to--leading order, $O(p^4)$, many new couplings enter. 
In the octet sector alone, there are 22 new low--energy 
constants \cite{EKW93}
in addition to the ones we have already encountered. The suspicion seems
well--founded that such an approach cannot have much predictive power. I 
will try to convince you that this suspicion is in general not justified.

Let us first turn to the dominant decay modes. It
turns out \cite{KMW91} that there are only seven combinations of coupling 
constants for the altogether\footnote{At lowest order, five of the
slope parameters vanish which explains the number seven mentioned before.} 
12 observables in $K\to 2\pi,\,3\pi$ decays. The resulting five 
relations \cite{KDHMW92} can be expressed as predictions for some of the
quadratic slope parameters in the $K\to 3\pi$ amplitudes. As shown in
Table \ref{tab:K3pi}, the agreement is very good for the two $I=1/2$
parameters where the data are most precise. The remaining predictions
for the $I=3/2$ slope parameters clearly need higher experimental precision
for a meaningful comparison. Thus, even for the dominant
nonleptonic $K$ decays there are predictions of the Standard Model that
remain to be tested.

\begin{table}
\caption{\it Predicted and measured values of the quadratic slope parameters
in $K \ra 3\pi$ amplitudes, all given in units of $10^{-8}$. The
table is taken from Kambor et al. \protect\cite{KDHMW92} and is based on the
CHPT calculation \protect\cite{KMW91} to $O(p^4)$.}
\label{tab:K3pi}
\vspace*{.5cm}
$$
\begin{tabular}{|c|c|c|} \hline
parameter & prediction & exp. value \\ \hline
$\zeta_1$ & $-0.47 \pm 0.18$ & $-0.47 \pm 0.15$ \\
$\xi_1$ & $-1.58 \pm 0.19$ & $-1.51 \pm 0.30$ \\
$\zeta_3$ & $-0.011 \pm 0.006$ & $-0.21 \pm 0.08$ \\
$\xi_3$ & $0.092 \pm 0.030$ & $-0.12 \pm 0.17$ \\
$\xi'_3$ & $-0.033 \pm 0.077$ & $-0.21 \pm 0.51$ \\ \hline
\end{tabular}
$$ 
\end{table}

All other nonleptonic $K$ decays are put into the category of rare decays.
The following classification takes into account the different structure
of chiral amplitudes for the various transitions. 

\subsection{Short--distance dominated transitions}
Here, CHPT cannot do much more than list the possible low--energy constants
(where all the short--distance structure resides) and estimate to
which extent the long--distance parts are suppressed. In addition
to the well--known decays \cite{Buchalla} $K\to \pi \nu \ol\nu$, the process 
$K_L\to \pi^+\pi^-\nu \ol\nu$ has recently been investigated \cite{GHLLV94}.
\subsection{Transitions with completely calculable $O(p^4)$ amplitudes}
In this group, none of the 22 low--energy constants occurring in general
at $O(p^4)$ actually appear in the amplitudes. There are still two
different cases to distinguish: either the amplitude vanishes altogether
at $O(p^4)$ or it does not.

Among the first transitions is $K_L\to \pi^0\pi^0\gamma$
where only an upper limit is available for the branching ratio \cite{
Barr94}. However, there is also $K_L\to \gamma\gamma$ which is
well--measured and yet cannot be counted as a success for CHPT. 
The amplitude for this decay vanishes at
$O(p^4)$ due to the Gell-Mann--Okubo mass formula for the meson masses,
but the corrections are large and not reliably calculable at present.
Although formally of $O(p^6)$, the actual size of the decay
amplitude as extracted from experiment is more like a typical $p^4$
amplitude. Unfortunately, this theoretical uncertainty influences also
the decay $K_L\to \mu^+\mu^-$ where the dispersive part of the two--photon
intermediate state cannot be reliably estimated \cite{Buchalla} for the 
same reason.  

Fortunately, $K_L\to \gamma\gamma$ is an exception rather than the rule
in nonleptonic $K$ decays. There are also transitions with non--zero $O(p^4)$ 
amplitudes which are completely calculable in terms of the leading--order
couplings $G_8$, $G_{27}$ appearing in loop amplitudes, among them
$K_S\to \gamma\gamma$, $K^0\to \pi^0 \gamma\gamma$ and
$K^0\to\pi^0\pi^0 \gamma\gamma$~.

\begin{figure}   
    \begin{center}
       \setlength{\unitlength}{1truecm}
       \begin{picture}(7.0,8.0)
       \put(-2.5,-5.0){\includegraphics{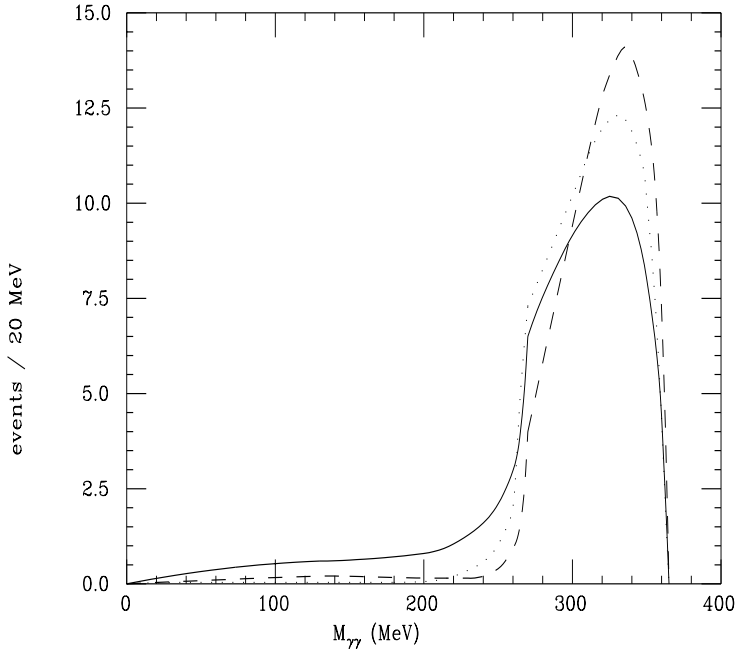}} 
       \end{picture}
    \end{center}
    \caption{\it Theoretical predictions for the $2\gamma$ invariant--mass
distribution in $K_L \rightarrow \pi^0 \gamma\gamma$. 
The dotted curve is the $O(p^4)$ contribution, 
the dashed and full curves correspond  
to the $O(p^6)$ calculation \protect\cite{CEP93}
without and with the appropriate
vector meson exchange contribution to reproduce the measured rate, 
respectively. The spectra are normalized to
the $50$ unambiguous events of NA31 (cf. Fig.~\protect\ref{fig:klpggexp}).}
    \protect\label{fig:klpggth}
\end{figure}

\begin{figure}   
    \begin{center}
       \setlength{\unitlength}{1truecm}
       \begin{picture}(7.0,8.0)
       \put(-2.0,-5.0){\includegraphics{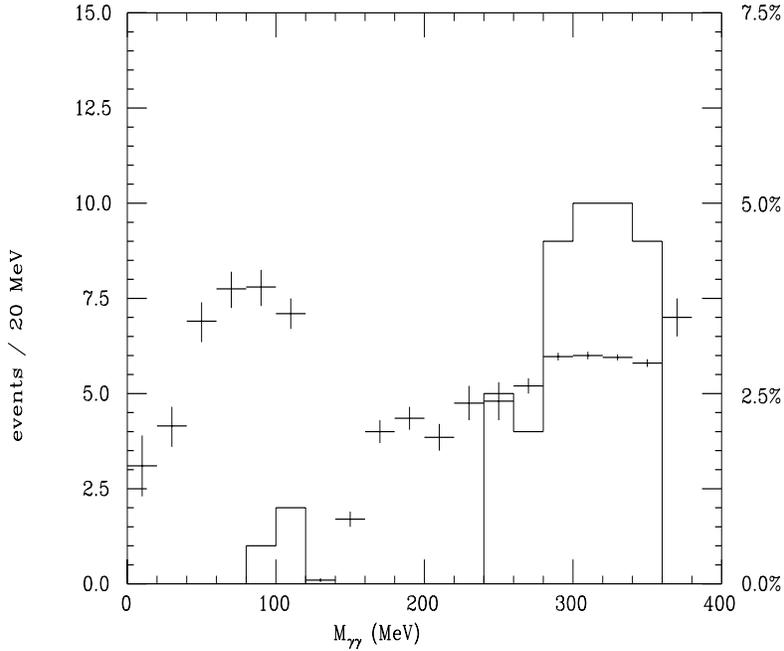}} 
       \end{picture}
    \end{center}
    \caption{\it $2\gamma$ invariant--mass distribution for 
unambiguous $K_L \rightarrow \pi^0 \gamma\gamma$ candidates from
NA31 \protect\cite{Barr92} (solid histogram).}
    \protect\label{fig:klpggexp}
\end{figure}

Let me briefly review the status of the two decays that have
already been measured. For $K_L\to \pi^0\gamma\gamma$, the experimental
branching ratios \cite{Barr92,Papa91}
\beq
BR(K_L \to \pi^0 \gamma \gamma)= \left\{ \ba{lr}
(1.7 \pm 0.2 \pm 0.2)\cdot{10}^{-6} & \mbox{ NA31}\\
(1.86\pm0.60\pm0.60)\cdot {10}^{-6} & \mbox{ E731}
\ea \right.
 \eeq
are substantially bigger than the chiral prediction \cite{EPR87b}
$BR \simeq 0.7\cdot 10^{-6}$. Higher--order corrections have been estimated
by several groups \cite{CDM93,CEP93,KH94}. The overall conclusion
is that the enhancement of the rate can be understood
but not really predicted by CHPT because of the uncertainties appearing
at $O(p^6)$ and higher. However, and this is really the main
message, the following two statements are then parameter--free predictions
of CHPT:
\bit\item
The two--photon mass spectrum can be predicted unambiguously once the rate
is fixed \cite{CEP93} and it is in perfect agreement with the NA31 
spectrum \cite{Barr92} as shown in Figs.~\ref{fig:klpggth} and 
\ref{fig:klpggexp}.

\item
When the same kind of analysis is applied to $K_S\to \gamma\gamma$,
there are essentially no corrections \cite{CEP93,KH94} of the type
occurring in $K_L\to\pi^0 \gamma\gamma$. Therefore, the CHPT 
prediction \cite{DEG86} of $O(p^4)$ for the branching ratio
\beq
BR(K_S\to\gamma\gamma)=2.0\cdot 10^{-6}
\eeq 
remains practically unchanged. The present agreement with the experimental
value \cite{Barr95}
\beq
BR(K_S\to\gamma\gamma)=(2.4\pm0.9)\cdot 10^{-6}
\eeq 
should therefore pass a more stringent test with better statistics.
The KLOE experiment at DA$\Phi$NE will certainly have enough statistics
for this purpose and is expected to improve the accuracy 
substantially \cite{Franzini}.
\eit

\subsection{Transition amplitudes with new couplings at $O(p^4)$}
By far the biggest group of nonleptonic kaon decays is characterized by
amplitudes where in addition to the already known constants $L_i$ of
Table \ref{tab:Li} some of the 22 (octet) couplings $N_i$ of the
nonleptonic weak Lagrangian of $O(p^4)$ appear. In Table \ref{tab:Ni},
a fairly complete list of such transitions and their dependence on the
nonleptonic low--energy constants is given. Without explaining the
seemingly obscure numbering of the $N_i$, it is easy
to see that only 9 of those constants enter
the various amplitudes: $N_{14}$,\dots,$N_{18}$ in so--called electric
amplitudes (without an $\ve$ tensor) and $N_{28}$,\dots,$N_{31}$ in
magnetic amplitudes (with an $\ve$ tensor). The four magnetic
constants are sensitive to the chiral anomaly \cite{BEP92,ENP94}.

This is not the place for a comprehensive discussion \cite{AEIN95} of the
transitions listed in Table \ref{tab:Ni}. Before discussing a few examples,
let me state the main conclusions:
\bit\item
All electric couplings $N_{14}$,\dots,$N_{18}$ can in principle be
determined phenomenologically.
\begin{table}
\caption{\it Decay modes to which the coupling constants $N_i$ contribute.
For the $3 \pi$ final states, only the single photon channels are
listed. For the neutral modes, the letters $L$ or $S$ in brackets
distinguish between $K_L$ and $K_S$ initial states in the limit of
CP conservation. $\gamma^*$ denotes a lepton pair in the final state.
If a decay mode appears more than once there are different Lorentz
structures in the amplitude.}
\label{tab:Ni}
\vspace*{.5cm}
$$
\begin{tabular}{|c|c|c|c|} \hline
$\pi$ & $2 \pi$ & $3 \pi$ & $N_i$ \\ \hline
$\pi^+ \gamma^*$ &$\pi^+ \pi^0 \gamma^*$ & & $N_{14}^r - N_{15}^r$\\
$\pi^0 \gamma^*~(S)$ & $\pi^0\pi^0\gamma^*~(L)$ & & $2 N_{14}^r + N_{15}^r$\\
$\pi^+ \gamma\gamma$ & $\pi^+\pi^0\gamma\gamma$ & & $N_{14} - N_{15}
-2 N_{18}$ \\
 & $\pi^+\pi^-\gamma\gamma~(S)$ & & " \\
 & $\pi^+\pi^0\gamma$ & $\pi^+\pi^+\pi^-\gamma$ & $N_{14}-N_{15}-N_{16}
-N_{17}$ \\
 & $\pi^+ \pi^- \gamma~(S)$ & $\pi^+\pi^0\pi^0\gamma$ & " \\
 & & $\pi^+\pi^-\pi^0\gamma~(L)$ & " \\
 & & $\pi^+\pi^-\pi^0\gamma~(S)$ & $7(N_{14}^r-N_{16}^r)+ 5(N_{15}^r
+ N_{17}^r)$ \\
 & $\pi^+\pi^-\gamma^*~(L)$ & & $ N_{14}^r - N_{15}^r -3(N_{16}^r
-N_{17})$\\
 & $\pi^+\pi^-\gamma^*~(S)$ & & $ N_{14}^r - N_{15}^r -3(N_{16}^r
+N_{17})$\\
 & $\pi^+\pi^0\gamma^*$ & & $ N_{14}^r + 2 N_{15}^r -3(N_{16}^r
-N_{17})$\\
\hline
 & $\pi^+\pi^-\gamma~(L)$ & $\pi^+\pi^-\pi^0\gamma~(S)$ & $N_{29} + N_{31}$ \\
 & & $\pi^+\pi^+\pi^-\gamma$ & " \\
 & $\pi^+\pi^0\gamma$ & $\pi^+\pi^0\pi^0\gamma$ & $3 N_{29} - N_{30}$ \\
 & & $\pi^+\pi^-\pi^0\gamma~(S)$ & $5 N_{29}  -N_{30}+ 2 N_{31} $ \\
 & & $\pi^+\pi^-\pi^0\gamma~(L)$ & $6 N_{28} + 3 N_{29} - 5 N_{30}$
\\ \hline
\end{tabular}
$$ 
\end{table}
\item
In contrast, only three combinations of the magnetic constants $N_{28}$,
\dots,$N_{31}$ appear in measurable decay amplitudes. Fortunately,
the theoretical expectations for these constants \cite{BEP92} are better 
founded than for the electric counterparts.
\item
A great number of interesting relations contained in Table \ref{tab:Ni}
remain to be tested. To the considered order in the low--energy expansion,
these relations are unambiguous predictions of the Standard Model
(low--energy theorems).
\eit

\subsubsection{$K^+\to \pi^+ l^+ l^-$}
The decay amplitude for this process ($l=e$ or $\mu$) depends, in addition
to explicitly known contributions, on the difference
$N_{14}-N_{15}$ (essentially the constant $w_+$ used previously \cite{EPR87a}).
Extracting this constant from the rate $\Gamma(K^+\to \pi^+ e^+ e^-)$, 
a two--fold ambiguity remains that was resolved by the 
measurement \cite{Alliegro92} of 
the spectrum in the invariant mass of the lepton pair. Once this constant
is determined, both rate and spectrum for the decay in the muon channel
are completely specified. The preliminary value 
$BR(K^+\to \pi^+ \mu^+ \mu^-)$=$(5.0\pm 0.4\pm 0.6)\cdot 10^{-8}$ reported
by the BNL-787 Collaboration at the Orsay Workshop \cite{Shinkawa}
is in excellent agreement with the theoretical prediction
$BR(K^+\to \pi^+ \mu^+ \mu^-)$=$(6.2^{+0.8}_{-0.6})\cdot 10^{-8}$.
The second prediction \cite{EPR87a} remains to be tested: unlike for the 
electron channel, the invariant--mass distribution of the muon pairs 
should be indistinguishable from phase space.
\subsubsection{$K^+\to \pi^+ \gamma\gamma$}
\begin{figure}   
    \begin{center}
       \setlength{\unitlength}{1truecm}
       \begin{picture}(5.0,7.3)
       \put(-3.0,-5.4){\includegraphics{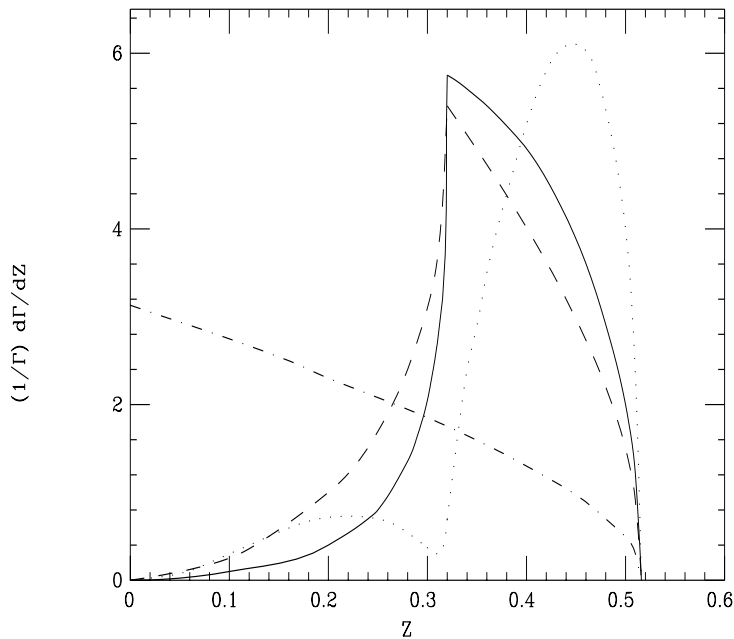}} 
       \end{picture}
    \end{center}
    \caption{\it Normalized distribution for the two--photon invariant
mass squared ($z=M^2_{\gamma\gamma}/M_K^2$) in 
$K^+ \to\pi^+ \gamma \gamma$ from CHPT \protect\cite{EPR88}
for several values of $\hat{c}$: $\hat{c}=0$ (full curve), $\hat{c}=4$ 
(dashed curve) and $\hat{c}=-4$ (dotted curve). The dash--dotted curve is 
the phase space distribution.}
    \protect\label{fig:Kpgg}
\end{figure}

Although this decay shares many features with $K_L\to \pi^0 \gamma\gamma$,
Table \ref{tab:Ni} shows that it depends on an unknown combination of
low--energy constants that is moreover different from the previous case of
$K^+\to \pi^+ l^+ l^-$. The combination $N_{14}-N_{15}-2 N_{18}$ is
related to the constant $\hat c$ introduced originally \cite{
EPR88}. For a reasonable range of values of this constant,
the spectrum in the two--photon invariant mass squared 
shown in Fig.~\ref{fig:Kpgg} has a very characteristic 
shape \cite{EPR88} similar to $K_L\to \pi^0 \gamma\gamma$. Preliminary
results from the BNL-E787 Collaboration \cite{Shinkawa,Nakano} are
consistent with this prediction. Of course, the rate is correlated with
the spectrum depending on the same constant $\hat c$.
A recent estimate of higher--order corrections \cite{AP96a} along similar
lines as for $K_L\to \pi^0 \gamma\gamma$ suggests an increase of the
rate by some 30 to 40 $\%$ over the value of $O(p^4)$.

\subsubsection{$K\to 3 \pi \gamma$}
There are four different modes in this channel only two of which
(for the charged kaon) have so far been observed experimentally. The full
calculation to $O(p^4)$ has just been completed \cite{AEIN96b}. To take
full advantage of the available information on the nonradiative transitions,
it is useful to generalize \cite{AEIN96a} the concept of Bremsstrahlung.
This is certainly the case for $K\to 3 \pi\gamma$, but it will also
be very useful in other reactions with four particles plus a photon. In
the present case, it turns out \cite{AEIN96b} that generalized 
Bremsstrahlung is an extremely good approximation\footnote{Less so for
$K_S\to \pi^+ \pi^-\pi^0\gamma$ where the nonradiative amplitude is
suppressed at lowest order.} to the amplitude of $O(p^4)$. The effect
of other contributions such as the coupling constants given in
Table \ref{tab:Ni} are completely hidden in the present experimental errors
for the nonradiative amplitudes. Much more precision would be needed
to be sensitive to those other contributions. As always, this can also
be phrased in a different, more positive way: the rates and
spectra of such processes are precisely predicted by the Standard Model 
in terms of the $K\to 3\pi$ parameters.

\section{Outlook}
The main success of CHPT in the field of kaon physics has been the
unified treatment of {\bf all} decay channels within the same framework
and the direct connection to the underlying Standard Model. For
semileptonic decays, the theory is in excellent shape. As the data improve,
some of the low--energy constants $L_i$ will become even better known, but
already now we have a very predictive scheme where to
$O(p^4)$ all parameters are known with reasonable accuracy.

We are still far from this state of affairs in the
nonleptonic sector. From the theoretical point of view, we need a better
understanding not only of the values of the low--energy constants $N_i$,
but also of their origin (as is the case \cite{EGPR89} for the $L_i$). 
There are several
attempts in this direction: $1/N_c$ expansion, lattice gauge theories, 
sum rules, chirally inspired models, \dots~. Even
with the limited knowledge we have of those constants, 
CHPT has been quite successful in the comparison with experiment
also in the nonleptonic sector.

However, most importantly of all, we need more
data to test the existing predictions of CHPT for $K$ decays. The
experimental program is well under way as we heard again during this meeting.
Allow me to close this talk with another plea to our experimental
colleagues: when you are out there searching for the few ``gold--plated"
events (exotic channels, CP violation, \dots), please do not neglect
the many ``standard" events. Many interesting tests of the Standard Model 
in $K$ decays are still ahead of us.

\section*{Acknowledgements}
I want to thank Konrad Kleinknecht for the invitation
and Lutz K\"opke for the efficient organization of this meeting in
such beautiful surroundings. This work has been supported in part by 
FWF (Austria), Project No. P09505--PHY and by HCM, EEC--Contract 
No. CHRX--CT920026 (EURODA$\Phi$NE). 
\end{document}